\documentstyle[aps]{revtex}

\def\prb{Phys. Rev. B}
\def\prl{Phys. Rev. Lett.}

\def\pr{Phys. Rev.}
\def\be{\begin{equation}}
\def\ee{\end{equation}}
\def\ba{\begin{eqnarray}}
\def\ea{\end{eqnarray}}

\def\C60{A$_x$C$_{60}$}

\begin{document}

\twocolumn[\hsize\textwidth\columnwidth\hsize\csname@twocolumnfalse\endcsname

\title
{Classification and Stability of Phases of the Multicomponent 
One-Dimensional Electron Gas}

\author{V.~J.~Emery$^2$, S.~A.~Kivelson$^1$, and O.~Zachar$^3$,  }
\address
{1)  Dept. of Physics,
U.C.L.A.,
Los Angeles, CA  90095}
\address{
2) Dept. of Physics,
Brookhaven National Laboratory,
Upton, NY  11973-5000}
\address{
3)  Laboratoire de Physique des Solides, U. Paris-Sud, 91405 Orsay , France}
\date{\today}
\maketitle 

\begin{abstract}

The classification of the ground-state phases of complex one-dimensional
electronic systems is considered in the context of a fixed-point strategy.
Examples are multichain Hubbard models, the Kondo-Heisenberg 
model, and the one-dimensional electron gas in an active environment. It 
is shown that, in order to characterize the low-energy physics, 
it is necessary to analyze the perturbative stability of the possible fixed points, 
to identify all discrete broken symmetries, and to specify the quantum
numbers and elementary wave vectors of the gapless excitations. Many  
previously-proposed exotic phases of multichain Hubbard models are shown 
to be unstable because of the ``spin-gap proximity effect.'' A useful tool 
in this analysis is a new generalization of Luttinger's theorem, which shows 
that there is a gapless even-charge mode in any incommensurate N-component 
system. 

\

\

\end{abstract}

]

\section{Introduction}

The basic theory of the low energy physics of the interacting one
dimensional electron gas (1DEG), both with and without spin, has been well
established for two decades. The purpose of this paper is to extend this
general analysis to obtain a classification of the stable fixed points of 
{\it multicomponent} one-dimensional electronic systems. Examples of such
problems include one-dimensional metals with several bands crossing the
Fermi surface, such as multichain Hubbard ladders,\cite
{multichain,2leg,3leg,white} and the ``1DEG in an active environment,''\cite
{ekz} of which the most studied example is the Kondo-Heisenberg model,\cite
{zachar-KLL,affleck} {\it i.e.}
a 1DEG interacting with a periodic array of localized
spins. While these models are still one dimensional, and are amenable to the
same methods of solution as the 1DEG, their added richness brings in
significant new physics. In particular, in the context of the theory 
of high temperature superconductivity, this class of models includes
some in which
a spin gap and a strongly divergent superconducting susceptibility 
derive from purely repulsive interactions.  Moreover, in these cases,
the driving force for the superconductivity is a lowering of the 
kinetic energy.\cite{ekz,pwa}

In 1D, even at zero temperature, states with a broken continuous symmetry
are destabilized by quantum fluctuations. However, there are states with
quasi-long-range order which can be characterized by the existence of
``quasi Goldstone modes'', {\it i.e.} gapless collective modes of the system
with a sound-like spectrum. The canonical example of a quasi Goldstone
mode is the longitudinal sound mode of a harmonic chain.\cite{kyoto}  For the
simple 1DEG, the relevant continuous symmetries are spin rotation invariance
(SU(2)), global gauge invariance (U(1)) associated with charge conservation,
and Galilean invariance. The latter is not an exact symmetry for the
electron gas on a lattice but, so long as the electron density is
incommensurate with the crystal, the low energy dynamics possess an exact
translational (chiral) symmetry.

\subsection{Classification of phases}

It is a remarkable feature of 1D that all the properties of such systems,
including fermionic correlation functions, can be expressed in terms of
bosonic fields (bosonization) corresponding to the quasi Goldstone modes.
Thus, it is possible to classify all thermodynamically-distinct ground-state
phases of any multicomponent 1DEG by specifying: (1) any spontaneously
broken discrete symmetries, such as the lattice translation symmetry or parity, 
and (2) the number and quantum numbers of the fundamental gapless modes.
The minimal quantum numbers of the gapless modes are charge, spin, and
(crystal) momentum. Our convention will be to focus on spin and charge modes
with the smallest non-zero momentum. Here ``spin modes'' have spin-1 and
charge 0, and ``charge modes'' have spin 0 and charge $2me$, where $m$ is an
integer. Simply counting gapless modes is insufficient; for instance, a
state with one gapless charge and one spin mode with the {\it same} momentum
(which we label $[c,s;2k_F]$ or $[cs]$ for short) is distinct from the state
in which they have different momenta, $[c;2k_F^{(1)}][s;2k_F^{(2)}]$ (or 
$[c][s]$ for short).

This scheme differs from the traditional method\cite{1Drev} of
classifying phases of the 1DEG in terms of the most divergent
susceptibilities, which is appropriate when the goal is to understand the
properties of {\it quasi}-one dimensional systems, since in most cases, weak
higher dimensional couplings will stabilize a true broken symmetry state
with the corresponding order at finite temperature. However, in the context
of the 1DEG {\it per se}, the Luttinger exponents, and hence the exponents
governing the divergence as $T\rightarrow 0$ of the various
susceptibilities, vary continuously with parameters. For example, one can
pass from a region in which the superconducting susceptibility is the most
divergent to one in which the CDW susceptibility is the most divergent
without encountering any thermodynamic singularities. Thus, {\it in the
strictly one-dimensional context}, the present classification is more
appropriate and, in this respect, it extends and corrects the
ground-breaking work of Lin, Balents and Fisher\cite{balents} on this
subject, while expanding on our earlier work on the general problem of a
``1DEG in an active environment.''\cite{ekz}

\subsection{Fixed-point strategy}

The concept of a fixed point of the renormalization group equations of a
field-theory Hamiltonian was introduced by Wilson\cite{wilson} for the
study of critical phenomena. This idea ultimately made its way into
many-body theory, where the renormalization group had been used for some
time. The main point is that the low-energy, long-distance physics of a
given model is controlled by the properties of the relevant stable fixed point or
critical fixed point of the renormalization group flows. A particularly
effective way of determining this behavior is to identify an
exactly-solvable field-theory model that starts in the neighborhood of an
unstable fixed point and flows to the same fixed point as the model in
question. This strategy justifies the use of the ``Toulouse limit'' to solve
the single-channel Kondo problem,\cite{alten} and of field-theory solutions of
the 1DEG with attractive backward scattering or umklapp scattering.\cite
{1Drev} It is important to note that a field-theory model that does not
exhibit spin-rotation invariance may flow to a spin-rotation invariant fixed
point Hamiltonian. The flexibility in the choice of solvable models allowed
by this behavior is frequently exploited, and it will be used later in this
paper.

A fixed-point strategy is more difficult to implement when there are many
degrees of freedom, as in multicomponent systems. There may be several
stable or critical fixed points, and it is necessary to carry out a ``global
renormalization''\cite{global} in order to determine which one controls the
low-energy physics of a given model. Usually, such a procedure must be
carried out numerically. Also, it is necessary to do a different stability
analysis for each fixed point separately.  The scaling dimension of 
any given operator is generally different at different fixed points, so
an interaction that is irrelevant
at one fixed point may become relevant at another.

One of our major findings is that ``the spin-gap proximity effect''\cite{ekz} 
is a powerful force for {\it destabiliizing} many putative fixed points of
multicomponent one-dimensional systems, and enhancing superconducting
correlations. The physics, which is driven by singlet pair tunnelling, is
analogous to the proximity effect in superconductivity. It serves to lock
the superconducting phases of two subsystems, and either generates a spin
gap in both subsystems, or transfers a spin gap from one subsystem to
another. We have proposed this effect as a mechanism of {high temperature
superconductivity}. A more detailed discussion is given in Sec. III.

\subsection{Outline of Paper}

The paper is organized as follows: In Sec. II the theory of the 1DEG is
reviewed from a statistical mechanical point of view, and the perspective
required for the analysis of the general problem is developed. In Sec. III
it is shown how this analysis can be extended to multicomponent systems.
Specifically, the classification of the possible fixed points in terms of
their spectrum and broken symmetries is discussed, and the conditions for
perturbative stability are derived. This section also contains a discussion
of generalizations of Luttinger's theorem to 1D, including a review of a
recent proof\cite{theorem} of the existence of a gapless neutral collective
excitation with momentum $2k_F^*$, and a new theorem concerning the
existence of a gapless charge $2me$ mode, 
where $m$ is an appropriate integer. The classification of phases and the
proof of the new theorem makes use of the continuum representation of the
so-called $\eta$-pairing operator, which is a product of fermion creation
operators with the {\it same} momentum.\cite{yang} 

Readers who are familiar with the theory of the 1DEG and are primarily 
interested in the illustrative examples may prefer to proceed directly to 
Sec. IV, where this general scheme is applied to the analysis of the global
phase diagram of the multichain Hubbard model and the Kondo-Heisenberg
model. Specifically, it is shown that 1) many of the proposed
partially-gapped phases\cite{balents} of the multichain Hubbard ladder are
destabilized by the ``spin-gap proximity effect'';\cite{ekz} 
2) the charge ordered ``stripe'' structures
which have been observed\cite{white} in numerical studies of multichain $t-J$ 
models have a (possibly non trivial) relation to the value of $2k_F^*$
derived from the generalized Luttinger's theorem\cite{theorem}; and 3) there
are several, thermodynamically-distinct spin-gap phases of the
Kondo-Heisenberg model.\cite{zachar} 

\section{The one component 1DEG}

To begin with, we consider the (well understood) theory of the single
component 1DEG with spin from the perspective of its quasi Goldstone modes.
All known zero-temperature thermodynamic states of the 1DEG can be
identified by: (1) spontaneously-broken discrete symmetries, such as parity,
(2) whether or not there is a gapless spin and/or charge collective mode,
(3) the smallest non-zero wave vector at which these modes are gapless,
which in analogy with Fermi liquid theory is called $2k_{F}$, (Equivalently, 
$2k_{F}$ characterizes the long-distance oscillatory behavior of appropriate
correlation functions.) (4) the velocity, $u$, and ``Luttinger exponent,'' 
$K$, of each gapless mode.

\subsection{Definition of Terms}

The definition of the terms used above requires some discussion since, in
much previous work, the identification of modes is derived from a particular
calculational scheme rather than from general principles. Because of the
absence of spontaneously broken spin-rotational symmetry, it is possible
even in the thermodynamic limit to classify all states by their spin quantum
numbers. Thus, by a ``spin mode'', we mean \cite{spinmodes} a neutral
excitation with spin 1; the existence of a gapless spin excitation will
typically show up as an asymptotic power-law behavior of the spin-spin
correlation function, $\langle \vec S_{2k_{F}}(t)\cdot \vec S_{-2k_{F}}(0)
\rangle \sim |t|^{1-2\delta_{S}}$. (Here, $\delta_{S}$, is the scaling
dimension of this operator.) A charge excitation refers to excitations made
by adding a small number of electrons to the system. Typically, this means 
\cite{pair} an excitation with spin-0 and charge 2e produced by adding a
singlet pair of electrons with total momentum $2k_{F}$, which will show up
in the asymptotic form of the ``$\eta$-pairing'' operator,\cite{yang}
defined below. It is implicit in the above classification scheme that a
state without a gapless spin mode has a ``spin gap'', and that a state
without a gapless charge mode has a ``charge gap''.

The physics of $2k_{F}$ is central to the following considerations. A
remarkable theorem of Yamanaka {\it et al},\cite{theorem} which we refer to
as the ``generalized Luttinger's theorem,'' fixes $2k_{F}=2\pi n/g$ at the
same value as it would have in a non-interacting electron gas; here $n$ is
the total electron density and $g=2$ because electrons have spin
1/2. The theorem, which we will discuss in more detail in Sec. II, implies
the existence of a gapless neutral excitation with momentum $2k_F$ as long
as the 1DEG is not commensurately locked to the underlying crystalline
lattice. We also show that the $\eta$-pairing operator, with the exact same
value of $2k_F$, creates a gapless excitation in the same circumstances.
When there are both gapless spin and charge modes, $k_{F}$ can also be
identified as the location of a non-analyticity in the single-particle
occupation probability, $<n_{k}>$, but when there is a spin or charge gap,
there is no sharp structure in the single-particle spectral function at all.
Thus, one must be careful in thinking of $k_F$ as a Fermi momentum.

Throughout this paper we will distinguish between the ``excitations'' and
the ``modes'' of the system. By a gapless ``excitation'' with given quantum
numbers we mean a set of excited states with energies which approach that of
the ground state in the thermodynamic limit. A ``mode'' refers to an
elementary excitation, which therefore has a well defined dispersion
relation. For example, above, we talked about the spin-1 charge-0
excitations of the system, although, in one dimension, under a broad range
of circumstances, the elementary excitations are, in fact, ``spinons'' with
spin-1/2 and charge 0. What this means is that the spin dynamic structure
factor will exhibit a branch cut, corresponding to a continuum of two spinon
excitations, rather than a pole corresponding to a magnon mode. When
classifying states in terms of gapless excitations, we have chosen not to
distinguish which are elementary and which are multi-particle excitations.

With this distinction in mind, the definition of the collective mode
velocities is obvious. The existence of Luttinger exponents, $K_{s}$ and 
$K_{c}$, is one of the triumphs of bosonization; they dictate the relation
between correlation functions expressed in terms of the original electronic
variables, and eigenmodes of the underlying bosonic free-field theory. For
any given lattice model, the quantum critical exponents $K_{\alpha}$ must be
determined by carrying out a global renormalization\cite{global} to the
appropriate fixed point and matching to the continuum theory. In general,
this procedure must be implemented numerically, either by studying the
long-distance behavior of correlation functions, or by studying the
finite-size scaling behavior of energy levels.

Two of the physically most important low energy fluctuations of the 1DEG are
the $2k_F$ CDW fluctuations and the zero-momentum (BCS-like) pairing
fluctuations.  CDW fluctuations are neutral and spinless. 
For example, for the repulsive $U$ Hubbard model with a
half-filled band, the $2k_{F}$ density-density correlation function has the
same asymptotic (power-law) form as the $2k_{F}$ spin-spin correlation
function, although there is manifestly a charge gap.\cite{korepin,stuff}

\subsection{Formal Implementation (Bosonization)}

Formally, the above discussion is equivalent to the statement that the
low-energy properties of the 1DEG are directly related to the properties of
two independent bosonic field theories with Hamiltonian densities 
\begin{equation}
{\cal H}_{\alpha }=\frac{u_{\alpha }}{2}\left[ (\partial _{x}\theta _{\alpha
})^{2}+(\partial _{x}\phi _{\alpha })^{2}\right] +V_{\alpha }\cos (\beta
_{\alpha }\phi _{\alpha })  \label{eq:sine}
\end{equation}
where $\alpha =c,s$ for the charge and spin fields, respectively, $\theta
_{\alpha }$ is the dual field to $\phi _{\alpha }$, or equivalently 
$\partial _{x}\theta _{\alpha }$ is the momentum conjugate to $\phi _{\alpha
} $. $V_{\alpha }$ can be set equal to zero\cite{su2} at the gapless fixed
point. Otherwise, when $V_{\alpha }$ is non-zero and relevant, {\it i.e.}
when $\beta _{\alpha }<\sqrt{8\pi }$, it sets the scale of the gap,\cite{gap}
according to the scaling relation $\Delta _{\alpha }\sim V_{\alpha }(\Lambda
/V_{\alpha })^{1-\beta _{\alpha }^{2}/8\pi }$ where $\Lambda $ is an
ultraviolet cutoff parameter. The Luttinger exponents, $K_{\alpha }$,
determine the value of $\beta _{\alpha }$ to be $\sqrt{8\pi K_{\alpha }}$,
and also specify the relationship between correlation functions expressed in
terms of the bosonic field operators, and physical correlation functions,
expressed in terms of the original electronic field operators, 
\begin{equation}
\psi _{\lambda ,\sigma }(x)={\cal N_{\sigma }}\exp \left[ i\lambda
k_{F}x-i\Phi _{\lambda ,\sigma }(x)\right]  \label{eq:bose}
\end{equation}
where ${\cal N_{\sigma }}$ contains both a normalization factor (which depends 
on the ultraviolet cutoff) and a ``Klein'' factor \cite{1Drev} (which can be 
implemented in many ways) so that $\psi _{\lambda ,\sigma }(x)$ anticomutes 
with ${\cal N_{\sigma ^{\prime }}}$ for $\sigma \neq {\sigma ^{\prime }}$
and commutes with it for $\sigma ={\sigma ^{\prime }}$. Also
\begin{equation}
\Phi _{\lambda ,\sigma }=\sqrt{\pi /2}\left[ (\tilde{\theta}_{c}+\lambda 
\tilde{\phi}_{c})+\sigma (\tilde{\theta}_{s}+\lambda \tilde{\phi}%
_{s})\right] ,  \label{eq:Phi}
\end{equation}
where $\lambda =\pm 1$ refers to left and right moving electrons, and 
$\sigma =\pm 1$ refers to the spin polarization. In Eq. (\ref{eq:Phi}), we
have expressed the fermion operators in terms of ``bare'' bosonic fields, 
$\tilde{\phi}_{\alpha }$, which are related to the interaction shifted normal
fields \cite{1Drev} that appear in Eq. (\ref{eq:sine}) by the canonical
(Bogoliubov) transformation, 
\begin{equation}
{\tilde{\phi}}_{\alpha }=\phi _{\alpha }\sqrt{K_{\alpha }}\ \ ;\ \ \tilde{%
\theta}_{\alpha }=\theta _{\alpha }/\sqrt{K_{\alpha }}.
\end{equation}
This transformation brings the Hamiltonian into canonical form, so that the
Luttinger exponents appear only in the relation between the fermionic and
bosonic fields, and implicitly in the values of $\beta _{\alpha }$.

>From Eq. (\ref{eq:bose}), it is a straightforward (and standard\cite{1Drev})
exercise to obtain bosonic representations of all interesting electron
bilinear and quartic operators. Physically, $\phi_c$ and $\phi_s$ are,
respectively, the phases of the $2k_F$ CDW and SDW fluctuations, and $%
\theta_c$ is the superconducting phase. 
The long-wavelength components of the charge ($\rho$) and spin ($S_z$)
densities are given by 
\begin{eqnarray}
\rho(x)=&&
\sum_{\lambda,\sigma}\psi^{\dagger}_{\lambda,\sigma}\psi_{\lambda,%
\sigma}=2k_F/\pi+ \sqrt{2K_c/\pi}(\partial_x \phi_c)  \nonumber \\
S_z(x)=&&
(1/2)\sum_{\lambda,\sigma}\sigma\psi^{\dagger}_{\lambda,\sigma}\psi_{%
\lambda,\sigma} =\sqrt{2K_s/\pi}(\partial_x \phi_s).  \nonumber
\end{eqnarray}

We also explicitly bosonize the $\eta $-pairing operator,\cite{yang} whose
correlations are sensitive to the presence or absence of a charge gap, 
\begin{eqnarray}
\eta _{\lambda } &=&\psi _{\lambda ,\uparrow }^{\dagger }\psi _{\lambda
,\downarrow }^{\dagger }  \label{eq:eta} \\
&\sim &\exp [i\sqrt{2\pi /K_{c}}(\theta _{c}+K_{c}\phi _{c})+2ik_{F}x]. 
\nonumber
\end{eqnarray}
This operator is not usually studied as its scaling dimension $\delta _{\eta
}=(K_{c}+K_{c}^{-1})/2$ is greater than 1 for $K_{c}>0$, so the
corresponding susceptibility is never divergent. However, it is interesting
in the present context as it has finite momentum and is independent of the
spin fields. It is easy to see from Eq. (\ref{eq:bose}) that spatial
translation by $x_{0}$ is equivalent to the chiral transformation $\phi
_{c}\rightarrow \phi _{c}-k_{F}x_{0}\sqrt{2/\pi K_{c}}$, which must be a
symmetry of the Hamiltonian for an incommensurate system. Similarly, gauge
invariance implies that the Hamiltonian is invariant under $\theta
_{c}\rightarrow \theta _{c}+{\rm const}$. As a consequence, the 
Hamiltonian must depend only on derivatives of $\theta_{c}$ and
$\phi_{c}$, so the $\eta $
operator defined above must always create a gapless excitation.

A final comment is in order at this point. The abelian representation
favored in the present paper is not manifestly spin-rotationally invariant.
This is an advantage whenever spin-rotational symmetry is broken at the
Hamiltonian level, as there is no need for special treatment of
symmetry-breaking terms. Spin-rotational invariance implies a specific value
of the Luttinger exponent $K_s$, which may be obtained by comparing the
spin-spin correlation functions for different spin directions. For example,
where there is no spin gap, spin-rotation invariance can easily be seen to
imply $K_s=1$ at the fixed point. (Slow flows as $K_s$ approaches 1 also can
give logarithmic corrections to various correlation functions). So far as we
know, in all cases studied to date, the fixed point value of $K_s$ in a
spin-gap phase is $K_s=1/4$ (or in other words, $\beta_s=\sqrt{2\pi}$), at 
which point the spin correlations are asymptotically equivalent
to those of a dimerized spin-1/2 Heisenberg model, but it is
conceivable that other discrete values could occur in other circumstances.

\section{The multicomponent 1DEG}

This section begins with a formal bosonized description of the
multicomponent 1DEG, continues with a discussion of the generalized
Luttinger theorems for this problem, and concludes with a detailed analysis
of the specific example of the two-component 1DEG. In particular, in this
latter part, it will be shown how the perturbative stability of each
potential fixed point Hamiltonian can be assessed.

\subsection{Bosonizing the multicomponent system}

First consider a system of N distinct 1DEG's, which may be bosonized as in
Eq. (\ref{eq:bose}): 
\begin{eqnarray}
&&\psi _{b,\lambda ,\sigma }(x)={\cal N}_{\sigma }^{b}\exp \left[ i\lambda
k_{F}^{b}x+i\Phi _{\lambda ,\sigma }^{b}(x)\right]  \label{eq:psi} \\
&&\Phi _{\lambda ,\sigma }^{(b)}=\sqrt{\pi /2}\left[ \tilde{\theta}%
_{c}^{(b)}+\lambda \tilde{\phi}_{c}^{(b)}\right] +\sigma \sqrt{\pi /2}\left[ 
\tilde{\theta}_{s}^{(b)}+\lambda \tilde{\phi}_{s}^{(b)}\right] ,  \nonumber
\end{eqnarray}
where $b=1$ to $N$ labels the different subsystems, the Klein factors $%
N_{\sigma }^{(b)}$ anticommute for $(a,\sigma )\neq (b,\sigma ^{\prime })$,
and the bosonic fields satisfy canonical commutation relations, 
\begin{equation}
\lbrack \tilde{\phi}_{\alpha }^{(a)}(y),\partial _{x}\tilde{\theta}_{\beta
}^{(a)}(x)]=i\delta _{a,b}\delta _{\alpha ,\beta }\delta (x-y).
\end{equation}
The tilde field variables appearing here are the bare fields, unshifted by
interactions.

In the continuum limit, the Hamiltonian density of this system consists of
three terms 
\begin{equation}
{\cal H}={\cal H}_c+{\cal H}_s+{\cal H}_{int}.
\end{equation}
Here ${\cal H}_c$ includes all the marginal interactions involving the
charge degrees of freedom, 
\begin{equation}
{\cal H}_c= \frac 1 2 \left\{ (\partial_x {\bf \tilde \theta}_c^T) {\bf W}_c
(\partial_x {\bf \tilde \theta}_c) + (\partial_x{\bf \tilde \phi}_c^T) {\bf V%
}_c(\partial_x{\bf \tilde \phi}_c) \right\},
\end{equation}
where ${\underline {\tilde \theta}}_c$ and ${\underline {\tilde \phi}}_c$
are column vectors with N components ${\tilde \theta}_{ci}$ and ${\tilde \phi%
}_{ci}$ respectively, and ${\bf W}_c$ and ${\bf V}_c$ are real, symmetric N $%
\times$ N matrices. So that the spectrum is bounded below it is necessary
and sufficient that the all eigenvalues of ${\bf W}_c$ and ${\bf V}_c$ be
positive. ${\cal H}_s$ is similarly defined for the spin degrees of freedom;
however, at the spin-rotationally invariant gapless fixed point, ${\bf W}_s=%
{\bf V}_s={\bf u}_s$, where $[{\bf u}_s]_{a,b}= u_{s}^{(b)} \delta_{a,b}$ is
the spin-velocity matrix. Finally, ${\cal H}_{int}$ contains the terms
nonlinear in the field variables (the various cosine interactions), which
when relevant lead to the opening of gaps in the spectrum, and when
irrelevant can be ignored.

For the case in which the nonlinear interactions are perturbatively
irrelevant, ${\cal H}$ is the fixed point Hamiltonian for a system with $N$
gapless charge and $N$ gapless spin modes. In this case, as in the
single-component problem, we perform a Bogoliubov transformation to normal
coordinates.\cite{mutta} A more detailed derivation is given in
Appendix A. First define the column vector ${\underline {\eta}}_i$ such that 
\begin{equation}
{\bf W}_c{\bf V}_c{\underline \eta}_i = u_{ci}^2 \hskip 0.1 cm {\underline %
\eta}_i  \label{eq:wv}
\end{equation}
where the $u_{ci}$ are the normal mode velocities, and 
\begin{equation}
{\underline \eta}_i^T {\bf W}_c^{-1} {\underline \eta}_j = \delta_{ij}.
\label{eq:vec}
\end{equation}
With these definitions, it is straightforward to show that 
\begin{equation}
{\bf W}_c = \sum_i {\underline \eta}_i \hskip 0.1 cm {\underline \eta}_i^T
\label{eq:w}
\end{equation}
and 
\begin{equation}
{\bf V}_c = {\bf W}_c^{-1} \sum_i u_{ci}^2 \hskip 0.1 cm {\underline \eta}_i %
\hskip 0.1 cm {\underline \eta}_i^T \hskip 0.1 cm {\bf W}_c^{-1}.
\label{eq:v}
\end{equation}
Then the Hamiltonian may be diagonalized by a canonical transformation to
new fields $\phi_i^{\prime}$ and their conjugate momenta $\partial_{x}
\theta_i^{\prime}$: 
\begin{eqnarray}
\phi_i = && u_i^{\frac{1 }{2}} {\underline \eta}_i^T {\bf W}_c^{-1} {%
\underline{\tilde \phi}}_c  \nonumber \\
\theta_i = && u_i^{-{\frac{1 }{2}}} {\underline \eta}_i^T {\underline {%
\tilde \theta}}_c .  \label{eq:bogo}
\end{eqnarray}
In transformed variables, the Hamiltonian consists of $N$ decoupled acoustic
normal modes, 
\begin{equation}
{\cal H}_{c}= \frac 1 2 \sum_i u_{ci} \left \{\partial_{x} \theta_{ci}^2 +
[\partial_x \phi_{ci}]^2 \right\}.
\end{equation}
The relation between the fermionic fields and the normal mode coordinates is
easily derived from this expression and Eq. (\ref{eq:psi}).

Finally, correlation functions of the untransformed fields can be expressed
in terms of the transformed fields using equation (\ref{eq:bogo}). A typical
operator has the form 
\begin{equation}
\hat O(x) \equiv \exp[ i\ ({\underline a}^T {\underline {\tilde \phi}}(x) + {%
\underline b}^T {\underline {\tilde \theta}}(x)) \ ]  \label{eq:operator}
\end{equation}
where ${\underline a}$ and ${\underline b}$ are N component real vectors,
and its zero-temperature equal-time correlation function is given by 
\begin{equation}
\langle \hat O(x) \hat O^{\dagger}(0) \rangle = [\Lambda |x|]^{-2\delta}
\end{equation}
where $\Lambda$ is an ultraviolet cutoff, and the scaling dimension 
\begin{equation}
\delta= \frac 1 {4\pi} \left[{\underline a}^T {\bf M}^{-1} {\underline a} + {%
\underline b}^T {\bf M} {\underline b}\right].  \label{eq:delta}
\end{equation}
where 
\begin{equation}
{\bf M} \equiv {\bf W}^{-{\frac{1 }{2}}} {\bf N} {\bf W}^{-{\frac{1 }{2}}}
\label{eq:m}
\end{equation}
with 
\begin{equation}
{\bf N}^2 \equiv {\bf W}^{\frac{1 }{2}} {\bf V} {\bf W}^{\frac{1 }{2}}.
\label{eq:n}
\end{equation}

The perturbative stability of the free-boson fixed point can be readily
analyzed by studying the scaling dimension of the various operators which
enter into ${\cal H}_{{int}}$. As usual, the stability of the fixed point
turns on whether there are any physically-allowed vectors ${\underline{a}}$
and $\underline{b}$ that lead to a scaling dimension less than 1, which
would imply that the operator is relevant. If any of these interactions is
relevant, it is necessary to identify the new fixed point to which the
system flows, and to study its properties. Typically, the effect of a
relevant interaction in ${\cal H}_{{int}}$ is to freeze out certain
fluctuations ({\it i.e.} to gap some modes) and at the same time produce a
renormalization of the matrices ${\bf V}$ and ${\bf W}$. This leads to a new
fixed point Hamiltonian, whose stability must be reassessed, since
operators that were irrelevant at the original fixed point could
become relevant at the new fixed point. This stability analysis will be
performed more explicitly in the two-component example discussed below.

\subsection{Generalization of Luttinger's Theorem}

The generalized Luttinger's theorem\cite{theorem} imposes an important
constraint on the allowed momenta at which gapless neutral
excitations occur. No matter how complex the system ({\it e.g.} no matter
how many bands cross the Fermi surface), unless there is an even integer
number of electrons per unit cell, there must be a zero energy excited state
with charge 0 and (for the case of zero net magnetization) with crystal
momentum $2k_{F}^{\ast }=\pi n_{T}$ where $n_{T}$ is the total electron
density, including all bands. Thus if a multicomponent system has gapless
modes at only one crystal momentum, it must be 
$2k_{F}^{\ast }$, and in a system with multiple values of $2k_{F}^{(b)}$,
there must be a set of integers $m_{b}$ such that 
\begin{equation}
2k_{F}^{\ast }=\sum_{b=1}^{N}m_{b}2k_{F}^{(b)}\ \ +\ \ {\rm reciprocal}\ \ 
{\rm lattice}\ \ {\rm vector}.
\end{equation}
(If some modes are charged, there is an obvious further constraint on the
integers implied by the neutrality of the composite mode at $2k_{F}^{\star }$%
.)

There is a second general constraint governing the existence of a gapless
charge excitation, which is discussed here for the first time.

This argument generalizes our earlier discussion of the $\eta $-pairing
mode. Consider the generalized $\eta $ operator, which creates 2N$_{c}$
right-moving electrons with spin 0: 
\begin{eqnarray}
\eta _{T,1} &=&\prod_{b=1}^{N_{c}}\psi _{b,1,\uparrow }^{\dagger }\psi
_{b,1,\downarrow }^{\dagger }  \label{eq:generaleta} \\
&=&(\ \ )\exp [i\sqrt{2\pi }(\tilde{\theta}_{c}+\tilde{\phi}%
_{c})+i2k_{F}^{\ast }x]  \nonumber
\end{eqnarray}
where $N_{c}$ is the number of ``extended'' charge modes and 
\begin{equation}
\tilde{\theta}_{c}=\sum_{b}\tilde{\theta}_{c}^{(b)}/\sqrt{N_{c}}\ \ {\rm and}%
\ \ \tilde{\phi}_{c}=\sum_{b}\tilde{\phi}_{c}^{(b)}/\sqrt{N_{c}}
\end{equation}
are the global superconducting phase and the dual CDW phase. Global gauge
invariance implies that the Hamiltonian is invariant under the transformation 
$\tilde{\theta}_{c}\rightarrow \tilde{\theta}_{c}+{\rm const.}$ Operationally, 
``extended'' charge modes are defined to be those modes
that acquire a non-zero phase under a global gauge transformation.
Similarly, spatial translation is equivalent to the phase shift, $\tilde{\phi%
}_{c}\rightarrow \tilde{\phi}_{c}+{\rm const.}$. As a consequence of these
invariances, ${\cal H}_{int}$ must dependent only on derivatives of $\tilde{%
\theta}_{c}$ and (so long as the system is incommensurate) on derivatives of 
$\tilde{\phi}_{c}$. Thus, the associated modes must be gapless. This implies
that $\eta _{T,1}$ (and of course, $\eta _{T,-1}$ as well) must create a
gapless, spin 0 charge $2N_{c}e$ excitation with crystal momentum $%
2k_{F}^{c} $, and that the $\eta $-correlations must fall like a
power-law with distance. In many cases, $2k_{F}^{c}=2k_{F}^{\ast }$ and $%
N_{c}=N$, the number of ``bands'' which cross the Fermi surface, but we will
encounter cases, such as the Kondo-Heisenberg model discussed in Section
IIIB, in which $N_{c}<N$.

This proof relies on the field-theoretic representation of operators; it is
desirable to generalize it to the actual lattice system, but we have not yet
succeeded in doing so.

Since a gapless, spinless, neutral excitation with momentum $2k_F^*$ always
exist on general grounds, it need not be listed when classifying phases.

\subsection{Classification of fixed points}

The essential steps in extending the above analysis 
are to identify the possible
fixed points of a multicomponent system, and then examine their perturbative
stability. 
As for the single-component system, {\it states are identified by their 
discrete broken symmetries and by the ``irreducible'' or minimal set of charge 
and spin carrying gapless excitations.} The gapless, spinless excitation at 
$2k_{F}^{*}$ implied by the generalized Luttinger's theorem may be left 
implicit. In contrast to the single-component 1DEG, it is necessary to
specify not only the modes but also their momenta (aside from 0), which are
no longer completely determined by the generalized Luttinger's theorem. For
instance, it will be seen that it is possible to encounter a state with a
gapless charge excitation at crystal momentum $2k_F$, and a gapless spin
mode at crystal momentum $2k_F^{\prime}\ne 2k_F$. Such a state will be
labelled $[2k_{F}:c][2k_F^{\prime}:s]$ or, leaving the values of the crystal
momentum implicit, $[c][s]$. In the canonical ordering to be adopted here,
the modes with the larger momentum, $2k_F > 2k_F^{\prime}$, will be listed
first. An interesting feature of the multicomponent system which is obviated
by the generalized Luttinger's theorem for the single component case is
that, when there are multiple values of $k_{F}$, their values can (and
generally will) shift continuously as a function of interactions.

Of course, it is always implicit that, if there exists a gapless excitation
at crystal momentum $2k_F$, then one can make gapless excitations with
integer multiples of $2k_F$, as well. However, it is clearly not sufficient
to specify the number of gapless modes, as proposed by Lin, Balents, and
Fisher. \cite{balents}
For example, the state
with gapless charge and spin modes with the same crystal momentum, $[cs]$,
is thermodynamically distinct from the states $[c][s]$ and $[s][c]$ in which
they occur at distinct crystal momenta.

\subsection{The two component 1DEG}

To make the discussion more concrete, and in particular to illustrate 
the nature of the stability analysis, we now consider the case of a 
two component 1DEG, $N=2$.
Two independent, decoupled, and generally inequivalent 1DEG's are separately
described by an appropriate fixed point (free boson) Hamiltonian. Clearly no
coupling between the two subsystems can be generated by any reasonable
renormalization group (RG) transformation. Thus the fixed points may be
specified for each subsystem separately. The discussion will be restricted
to the spin-rotationally invariant case, although this is easily generalized.

The next step is to determine the circumstances in which each fixed point is
stable with respect to weak interactions between the two 1DEG's. In general,
whenever a given fixed point is stable for some range of parameters, there 
is no more to say. If the fixed point is unstable, the character of the
stable fixed point to which the Hamiltonian flows under renormalization must
be determined. The new fixed point could, in principle, have only gapless
modes, although, usually some modes that were gapless become gapped.

If the two subsystems are mutually incommensurable, ($k_{F}^{(1)}/
k_{F}^{(2)}=$ irrational) the only potentially perturbatively
relevant couplings are those
that do not transfer momentum between the two systems. The most relevant
terms are quartic in fermion operators and are of three types. The
interaction piece of the Hamiltonian density is given by 
\begin{equation}
{\cal H}^{\prime}= {\cal H}^{\prime}_1 + {\cal H}^{\prime}_2 + {\cal H}%
^{\prime}_3
\end{equation}
where 
\begin{equation}
{\cal H}^{\prime}_1 = V\rho^{(1)}\rho^{(2)}+ V^{\prime} j^{(1)}j^{(2)}
\end{equation}
with 
\begin{eqnarray}
\rho^{(b)}(x)=&& \sum_{\lambda,\sigma} \psi_{b,\lambda,\sigma}^{\dagger}
\psi_{b,\lambda,\sigma} \\
j^{(b)}(x)=&& \sum_{\lambda,\sigma} \lambda\psi_{b,\lambda,\sigma}^{\dagger}
\psi_{b,\lambda,\sigma},
\end{eqnarray}
\begin{equation}
{\cal H}^{\prime}_2 = J\vec S^{(1)}\cdot \vec S^{(2)}+ J^{\prime}\vec j%
^{(1)}\cdot \vec j^{(2)}
\end{equation}
with 
\begin{eqnarray}
\vec S^{(b)}(x)=&& \sum_{\lambda,\sigma,\sigma^{\prime}}
\psi_{b,\lambda,\sigma}^{\dagger}\vec \tau_{\sigma,\sigma^{{\prime}}}
\psi_{b,\lambda,\sigma^{\prime}} \\
\vec j^{(b)}(x)=&& \sum_{\lambda,\sigma,\sigma^{\prime}}\lambda
\psi_{b,\lambda,\sigma}^{\dagger}\vec \tau_{\sigma,\sigma^{{\prime}}}
\psi_{b,\lambda,\sigma^{\prime}}
\end{eqnarray}
and 
\begin{equation}
{\cal H}^{\prime}_3 = \left [{\cal J}_{s} \Delta^{(1) \dagger}\Delta^{(2)} +%
{\rm H.c.}\right] +\left[{\cal J}_{t} \vec\Delta^{(1) \dagger}\cdot \vec%
\Delta^{(2)}+{\rm H. c.}\right],
\end{equation}
where 
\begin{eqnarray}
\Delta^{(b)}(x)=&&\sum_{\lambda} \psi_{b,\lambda,\uparrow}^{\dagger}
\psi_{b,-\lambda,\downarrow}^{\dagger} \\
\vec \Delta^{(b)}(x)=&& \sum_{\sigma,\sigma^{\prime}}
\psi_{b,1,\sigma}^{\dagger}\vec \tau_{\sigma,\sigma^{{\prime}}}
\psi_{b,-1,\sigma^{\prime}}^{\dagger}.
\end{eqnarray}
Here $\vec \tau$ are the Pauli matrices. Of these interactions, the charge
and current-density interactions in ${\cal H}^{\prime}_1$, are marginal, 
{\it i.e. } they are quadratic in boson variables, and so must (and can) be
absorbed into the definition of the fixed-point Hamiltonian density and
treated exactly, as in subsection A. The perturbative stability analysis is
then performed with respect to the remaining interactions, ${\cal H}%
^{\prime}_2$ and ${\cal H}^{\prime}_3$, by computing the scaling dimensions
of these operators, as in Eq. (\ref{eq:delta}).

This stability analysis was carried out previously for the two chain problem
in Refs. \cite{mutta} and \cite{ekz}. The results are algebraically
complicated, but are simplified, without significant loss of physical
insight, by considering systems in which 
\begin{equation}
V^{\prime }=-(v_{c}^{(2)}/v_{c}^{(1)})(K_{c}^{(1)}K_{c}^{(2)})V,
\end{equation}
where $K_{\alpha }^{(b)}$ and $v_{\alpha }^{(b)}$ are the Luttinger exponent
and velocity at the decoupled fixed point of subsystem $b=1$, $2$, with $%
\alpha =c$, $s$ for charge and spin modes, respectively. In this case, 
\begin{eqnarray}
\delta _{(\vec{S}_{1}\cdot \vec{S}_{2})} &=&\frac{1}{4}\left(
K_{s}^{(1)}+K_{s}^{(2)}+1/K_{s}^{(1)}+1/K_{s}^{(2)}\right)  \nonumber
\\
\delta _{(\vec{j}_{1}\cdot \vec{j}_{2})} &=&\delta _{(\vec{S}_{1}\cdot \vec{S%
}_{2})} \label{eq:dimensions} \\
\delta _{(\Delta _{1}\Delta _{2})} &=&\frac{1}{4}\left(
A/K_{c}^{(1)}+B/K_{c}^{(2)}+K_{s}^{(1)}+K_{s}^{(2)}\right)  \nonumber \\
\delta _{(\vec{\Delta}_{1}\cdot \vec{\Delta}_{2})} &=&\frac{1}{4}\left(
A/K_{c}^{(1)}+B/K_{c}^{(2)}+1/K_{s}^{(1)}+1/K_{s}^{(2)}\right) ,  \nonumber
\end{eqnarray}
with 
\begin{eqnarray}
A &=&\sqrt{1+\frac{4VV^{\prime }}{(\pi v_{c}^{(2)})^{2}}} \\
B &=&\left( 1-\frac{2VK_{c}^{(1)}}{\pi v_{c}^{(1)}}\right) ^{2}/\sqrt{1+%
\frac{4VV^{\prime }}{(\pi v_{c}^{(1)})^{2}}}.  \nonumber
\end{eqnarray}
The weak-coupling fixed point of two incommensurate 1DEG's is stable if
all of these dimensions are greater than 1, and is unstable otherwise. For
weak interactions, any gapless charge modes have $K_{c}^{(b)}$ near 1, and $%
K_{c}^{(b)}$ generally increases with increasingly strong repulsive
interactions. When the fixed point Hamiltonian is spin rotationally
invariant, $K_{sj}=1$.

The expressions in Eq. (35)
were derived for the gapless fixed
point Hamiltonian, but it is relatively easy to deduce how these scaling
dimensions are altered at a strong-coupling fixed point in which certain
fluctuations are frozen out by the presence of relevant interactions of the
form $\cos [\beta \phi _{c}^{(b)}]$ and/or $\cos [\beta \phi _{s}^{(b)}]$
which open a gap. This is equivalent to replacing $K_{c}^{(b)}$ by an
effective Luttinger exponent, $K_{c}^{(b)}\rightarrow 0$ and/or $%
K_{s}^{(b)}\rightarrow 0$. Conversely, if the fluctuations of the dual
phases are suppressed by a relevant interaction of the form $\cos [\beta
\theta _{c}^{(b)}]$ and/or $\cos [\beta \theta _{s}^{(b)}]$, these
expressions should be evaluated in the limit $K_{c}^{(b)}\rightarrow \infty $
and/or $K_{s}^{(b)}\rightarrow \infty $. Other types of strong-coupling
fixed points can be analyzed in the same fashion.

It is worth commenting, briefly, on the physical implications of the
dependence of these various scaling dimensions on the parameters in the
fixed-point Hamiltonian:

The spin interactions, ${\cal H}_{2}^{\prime }$, are manifestly unimportant (%
$\delta _{(\vec{S}_{1}\cdot \vec{S}_{2})}$ and $\delta _{(\vec{\Delta}%
_{1}\cdot \vec{\Delta}_{2})}$ are infinite) if either subsystem has a spin
gap. This makes good physical sense. If neither system has a spin gap, then
the constraints of spin-rotation invariance imply that these interactions
are marginal ($\delta _{(\vec{S}_{1}\cdot \vec{S}_{2})}=1$); further
analysis ({\it i.e.} carrying out the perturbative analysis to order $J^{2}$%
), following on the work of Sikkema {\it et al.} \cite{affleck} on the
Kondo-Heisenberg problem, shows that for antiferromagnetic couplings ($J>0$%
), these interactions are marginally relevant while for ferromagnetic
couplings ($J<0$), they are marginally irrelevant. Sikkema {\it et al.}
speculated that in the antiferromagnetic case, the system scales to a
strong-coupling fixed point with $J$ large and a total spin gap. This
conclusion is supported by numerical studies carried out by these same
authors, and by additional analytic work by one of us.\cite{zachar}

The singlet pair tunnelling interaction, ${\cal H}^{\prime}_3$, has its
scaling dimension significantly reduced if either or both subsystems have a
spin gap, since then the effective $K_{s}^{(b)}$ in Eq. (35) is zero. 
For instance, if subsystem $1$ has a spin gap, and subsystem $2$ does
not, then from Eq. (35)
\begin{equation}
\delta_{(\Delta_1 \Delta_2)}=\frac 1 4 \left( A/K_{c}^{(1)} +
B/K_{c}^{(2)}\right) +\frac 1 4 .  \label{eq:spingap}
\end{equation}
The underlying physics is analogous to the proximity effect in
superconductivity, and we have named it\cite{ekz} the ``spin gap proximity
effect:''  because subsystem $1$ has a spin gap, it  is already substantially
superconducting,\cite{rvb} so it can readily infect any coupled subsystem
with its superconducting character. From this point of view, one would
expect a relevant pair tunnelling interaction
to induce pairing correlations in subsystem $2$ ({\it i.e.} to open a
spin gap) and to lock the superconducting phases of the two subsystems ({\it 
i.e.} to gap the out-of-phase CDW mode).  We have confirmed
the validity of this expectation  by
exact an solution of this problem in a particular solvable limit.\cite{ekz} 
When ${\cal J}_{s}$ is relevant, {\it i.e.}
whenever $\delta_{(\Delta_1\cdot \Delta_2)} <1$,
the stable fixed-point behavior is characterized
by a total spin gap, and a locking of the charge degrees of freedom of the
two subsystems. Indeed, this effect is very efficient at destabilizing any
fixed point with a partial spin gap; for instance, any [cs][c] fixed point
is most often unstable for non-zero ${\cal J}_{s}$, and flows to the [c]
fixed point. The effect of a relevant triplet pair tunnelling interaction,
${\cal J}_t$, has not yet been thoroughly investigated.

When the two subsystems are mutually commensurate, or nearly commensurate,
the above stability analysis becomes more complicated. We defer
detailed discussion of this problem to a later date. However, a few
interesting features of the problem can be understood on the basis of very
general considerations. In the first place, the decoupled fixed point of two
1DEG's with the same values of $2k_F$ is always perturbatively unstable,
unless at least one system has a fully gapped spectrum. This follows directly
from the observation that in all the known phases of the 1DEG, at least one
susceptibility is enhanced relative to non-interacting 
electrons.\cite{1Drev} What this also implies is that,
if we start with two decoupled 1DEG's with $2k_F^{(1)}$ nearly equal to 
$2k_F^{(2)}$, and then gradually turn on interactions between them, there is
a strong tendency to induce transfer of electrons between the two
subsystems, with a cost of unperturbed energy but a gain of energy from the
relative commensurate locking of the two subsystems. As a result, one
expects a {\it relative} incommensurate to commensurate transition as a
function of increasing interaction strength in such systems. 
As for the 1DEG itself, the situation is somewhat more complicated for higher
order {\it relative} commensurabilities, since newly-allowed interactions 
are generally irrelevant when the interactions in the 1DEG are weak, and 
relevant only when they are sufficiently strong and sufficiently 
long-ranged.\cite{commens}

\section{Application to specific Model Problems}

To demonstrate the utility of this analysis, we conclude with a discussion
of four specific problems that have been of considerable recent theoretical
interest.

\subsection{Critique of the Perturbative RG Analysis of the N-chain Hubbard
model}

There have been a number of recent papers concerning the phases of the
N-chain Hubbard model, following the early work of Varma and 
Zawadowski.\cite{VZ} In particular, in two interesting papers, Balents,
Fisher and Lin\cite{balents} analyzed the renormalization group flows in
the neighborhood of the {\it non-interacting fixed point}, by computing the
beta function to lowest order in powers of $U/t$, where $U$ is the on-site
repulsion between electrons and $t$ is the intersite hopping matrix. (More
generally, they allowed for possibly different values of the hopping
amplitudes, $t$ and $t^{\prime }$, respectively parallel and perpendicular
to the chain direction.) Specifically, they identified which interactions
are perturbatively most relevant for various geometries of the chains, and
as a function of the electron concentration per site and the ratio $%
t^{\prime }/t$. They then conjectured a phase diagram by analyzing the
nature of the fixed point obtained by bosonizing the model with the relevant
interactions taken to infinity, and all others neglected.

This analysis has, we believe, three flaws, which lead to significant errors
in the resulting phase diagram and other conclusions. 1) The only rigorous
conclusion that can be drawn from a perturbative RG analysis when there are
relevant interactions is that the initial fixed point is unstable, and that
therefore the asymptotic physics is controlled by another fixed point.
(Tracing the effects of the perturbatively relevant interactions to strong
coupling is, {\it at best}, suggestive of the character of the new fixed
point.) 2) Even if we accepted the perturbative analysis of the nature of
the interactions which are important at the strong-coupling fixed point to
which the Hamiltonian flows, it is essential to perform a perturbative
stability analysis at the new conjectured strong-coupling fixed point to
make certain that it is, in fact stable. This was not done by Balents,
Fisher and Lin. Specifically, their analysis ignores the possibility that
interactions that are marginal or irrelevant at the weak-coupling fixed
point, can become relevant at the strong-coupling fixed point. 3) Because of
the presence of marginal interactions, {\it e.g.} interactions which shift
the various collective mode velocities, it is an incorrect procedure to
perturb about the {\it non-interacting} fixed point. The proper perturbative
RG analysis should include the effects of the marginal interactions {\it %
exactly}, as in Section IIIA, and should be performed about the appropriate
free boson fixed point. Since the various collective mode velocities enter
the one-loop RG equations, this makes small (but at times important)
differences in the character of the weak-coupling flows

\subsubsection{Stability of Phases:}

How do these general observations affect the conclusions of Lin, Balents,
and Fisher\cite{balents} concerning the phase diagram? Where they find no
relevant interactions in their approach, and conclude that all of the modes
of the non-interacting system remain ungapped ({\it e.g.} where they find
phases of the form [cs]$^{n}$, or CnSn in their notation), or where they
find completely gapped phases, or phases with only a single, ungapped charge
mode ({\it e.g.} phases labeled [c] or C1S0 in their notation) the only
differences between our analysis and theirs are ${\cal O} (U/t)$ shifts of
the locations of various phase boundaries due to the effect of marginal
interactions. However, all of the partially spin-gapped phases, (such as the
phase [cs][c], which, in their notation is C2S1), are destabilized by the
spin-gap proximity effect. To see this, note that the pair tunnelling
interaction, like all other interactions, has scaling dimension $\delta=1+%
{\cal O}(U/t)$ at the non-interacting fixed point; the opening of a spin
gap, with all other interactions held small (of order $U/t$) reduces this
dimension by a finite amount, {\it e.g.} to $\delta=3/4+{\cal O}(U/t)$ in
the example in Eq. (\ref{eq:spingap}). Thus, of all the conjectured phases
in their phase diagrams, only the familiar phases with one or fewer gapless
charge and spin modes, and the novel totally gapless phases, [cs]$^m$, are
stable at weak interactions.

Of course, as is implied by point 1) above, it cannot be ruled out that during 
the flow to strong coupling, interactions other than the ones identified in
perturbation theory will get large, and this could alter the stability of
the partially-gapped phases. However, we suspect that the partially-gapped
phases are not, generically, stabilized for large $U/t$, because strong
interactions can produce significant shifts in the values of $k_{F}^{(b)}$,
which then permits the commensurate locking of different subsystems; indeed,
this conclusion was reached, previously, by Schulz.\cite{schulz} In this
regard it is worth noting that there have been extensive numerical\cite
{numerical} studies of various N-leg Hubbard ladders for intermediate to
strong $U/t$, in none of which has evidence for these exotic phases been
reported.

\subsection{Example of the spin-gap proximity effect: the asymmetric 3-chain
Hubbard Model}

In Ref. (\cite{ekz}), where the spin-gap proximity effect was first
elucidated, the simplest model system to which it was applied was the
asymmetric three-leg Hubbard ladder. The predictions 
made there were later confirmed
in numerical experiments on the symmetric three leg $t-J$ ladder.
\cite{3leg} For the reasons outlined above, these
results are in clear disagreement with the predictions based on the
perturbative RG analysis of Lin, Balents, and Fisher.\cite{balents} We
briefly review the analysis here, as an illustrative example.

As in Ref. (\cite{ekz}), consider an asymmetric system, with a two-leg
Hubbard ladder weakly coupled to a one-leg Hubbard ``chain'', with a
difference in site energy $\epsilon $. For concreteness, take $U$ to be
large, and consider the phase diagram as a function of electron density,
although in Ref. (\cite{ekz}) it was constructed as a function of $\epsilon $.
The method of analysis, as presented above,
first neglects the coupling between the ladder and the chain, and then
assesses its effect on the final result.

When the electron density per site is $n=1$, the system is manifestly
insulating. The two-leg ladder, on its own, has a spin gap while the chain
has a gapless spinon mode. Because of the spin gap, this fixed point is
stable for weak chain-ladder couplings, so the phase is [s].

For $\epsilon > 0$, when the density of electrons is reduced slightly to $%
n=1-x$ with the number of ``doped holes'' $x << 1/3$, the added holes go
onto the chain. Because the interactions in the chain are repulsive and
spin-rotationally invariant, and the electron density is incommensurate, the
chain will form a Luttinger liquid state with gapless charge and spin modes.
Pair tunnelling between the chain and the ladder induces an effective
attraction between spinons, but because there is an energy denominator $%
2\epsilon^{*}$ (where $\epsilon^{*}$ is the renormalized energy to transfer
a singlet pair of electrons from the ladder to the chain), the bare
repulsion between electrons on the chain is the dominant interaction, so the
decoupled fixed point is stable; this phase is [cs].

Finally, with increasing doping, although still in the regime $x < 1/3$, the
value of $\epsilon^{*}$ decreases steadily due to the repulsion between
doped holes so, if $|\epsilon|$ is not too large, we reach a regime in which
the pair tunnelling between the chain and the ladder becomes significant.
Now, via the spin-gap proximity effect, the chain becomes infected with the
spin gap of the ladder. The result is a phase which has a total spin gap,
and only the charge 2e and neutral spinless modes implied by the generalized
Luttinger's theorem; this phase is [c].

The system studied numerically\cite{3leg} is, in fact, the three chain $t-J$
ladder. The differences between the $t-J$ and
Hubbard models for intermediate to strong $U$ are not believed to be very
significant in the present context. Because of the boundary conditions, the
central chain of the ladder is physically distinct from the two edge chains.
Even though the bare difference
in site energy $\epsilon =0$, there is manifestly a non-zero value of $%
\epsilon ^{\ast }$. 
Thus, there is, in fact,
a very close (although not entirely quantified) relation between the system
we analyzed theoretically, and that studied in the numerical experiments, so
it is not surprising that the reported phase diagrams agree. In the
numerical experiments, with $J/t\sim 0.5$, the critical value of doping at
which the transition from [cs] to [c] occurs is $x\approx 0.06$; the very
small value of this critical density reveals the robustness and strength of
the spin-gap proximity effect. The perturbative RG analysis of Lin, Balents,
and Fisher leads to a phase diagram in which the undoped system is
(correctly) in the [s] phase, but has the doped system exhibiting the
[s][cs] phase over the relevant range of $x$; this is a specific case of a
partially spin-gapped phase which, as we argued above, should be generally
unstable to the formation of a fully spin-gapped [c] phase due to
pair-tunnelling interactions.

\subsection{Concerning the $N=6$ $t-J$ Cylinder}

Recent important advances in the numerical evaluation of the ground-state
properties of correlated systems have allowed the study of much larger $t-J$
and Hubbard clusters than before. White and Scalapino,\cite{white}
considered 6-component $t-J$ systems with cylindrical boundary conditions, 
{\it i.e.} periodic boundary conditions in the finite direction and open
boundary conditions along the chains. To draw conclusions concerning 
the 2d $t-J$ model, it is necessary to perform a 
two-dimensional finite size scaling analysis of these results in order to 
extrapolate to the thermodynamic limit. So far, it has not been
possible to do so, and the conclusions of White and Scalapino 
disagree with those of other studies 
of comparably large systems \cite{manousakis} which did do a finite size 
scaling analysis. Consequently, it is still unclear to what extent
these results are representative of the actual ground state of the two-dimensional
system. However, we may imagine that the results of White and Scalapino are
representative of the ground-state properties of an infinite length 6-leg
cylinder, which itself is an example of a multicomponent one-dimensional 
system.

The principal finding of White and Scalapino (obtained for $J/t=0.35$) is
that the ground state exhibits ``stripe'' correlations in the expectation
value of the charge-density operator. For the 6-leg cylinder at small
density of doped holes, $x$, the period of the observed density oscillations
is $\lambda _{6}=2/3x$. (Here, units of length are chosen so that the
lattice constant is equal to 1.) Of course, since this is one dimension, the
density-wave order observed by White and Scalapino on finite length systems
should be interpreted as the period of power-law CDW correlations in the
infinite system. Now, the value of $2k_{F}^{\ast }=G+2\pi /\lambda ^{\ast }$%
, where $G$ is a reciprocal lattice vector, corresponds to density-wave
correlations with wavelength $\lambda _{6}^{\ast }=1/3x$, so the period
found by White and Scalapino is twice that required by the generalized
Luttinger's theorem. In other words, the fundamental gapless, spinless
neutral CDW mode of the system occurs at a wave number $(1/2)2\pi /\lambda
^{\ast }$, and the excitation at $2k_{F}^{\ast }$ is thus a second harmonic. 

It is also easy to see from the present analysis that these cylinders are
not good candidates for high temperature superconductors, since there are no
gapless charge 2e excitations of this system. 
At present, it is not clear to us whether the system supports gapless
excitations with charge $4e$, corresponding to the injection of an
additional ``stripe'' into the system, or whether because there is a
tendency for spin correlations to suffer a $\pi $ phase shift across a
stripe, it is necessary to inject charge $8e$ corresponding to a pair of
``stripes.'' 

\subsection{Kondo-Heisenberg array}

The Kondo-Heisenberg model is the simplest example of a metallic system
(here, a 1DEG) coupled to an insulating antiferromagnet (here, a spin 1/2
Heisenberg chain). The large (or infinite) charge gap in the spin chain
implies that charge-transfer interactions, such as pair tunnelling, are
unimportant. The dominant interactions involve the spin density, {\it i.e.}
they are the Kondo interaction between the conduction electron spin $\vec{s}%
\left( x\right) $ and localized ``impurity'' spins $\vec{\tau}_{j}$, 
\begin{equation}
H_{K}=J_{K}\sum_{j}{\bf \vec{\tau}}_{j}\cdot {\bf \vec{s}}\left(
x_{j}\right) ,
\end{equation}
where $x_{j}$ are the positions of the localized spins and the
Heisenberg interaction between nearest-neighbor localized spins, 
\begin{equation}
H_{0}^{Heis}=J_{H}\sum_{j}{\bf \vec{\tau}}_{j}\cdot {\bf \vec{\tau}}_{j+1}.
\end{equation}
Also it will be assumed that $J_{H}\ll E_{F}$, which is typically
true in physical applications. The resulting Kondo-Heisenberg Hamiltonian is 
\begin{equation}
H=H_{0}^{1DEG}+H_{0}^{Heis}+H_{K}
\end{equation}
where, the subscript $0$ refers to the Hamiltonian of the decoupled system
with 
\begin{equation}
H_{0}^{1DEG}=-iv_{F}\sum_{\sigma ,\lambda }\lambda \int dx\psi _{\lambda
,\sigma }^{\dagger }\partial _{x}\psi _{\lambda \sigma }.
\end{equation}
It will be assumed that the relative concentration of localized spins is $c=1/b<1$, 
{\it i.e.} $x_{j}=jb$, that the two systems are relatively incommensurate,
and that $2k_{F}$ is incommensurate with the underlying lattice. The
effective Fermi wave number (in the sense of the generalized Luttinger's
theorem) for the 1DEG and the spin chain are $2k_{F}$ and $2k_{F}^{Heis}=\pi
/b$ respectively, so for the coupled system, \cite{theorem} there must be
a gapless neutral excitation with wave number $2k_{F}^{\ast }=2k_{F}+\frac{%
\pi }{b}$, and a charge $2e$ excitation with momentum $2k_{F}$.

The determination of the phase diagram of this model provides a further
example of the application of the methods developed above. We shall give a
brief physical description of the origin of the various phases --- the
theoretical manipulations may be found in Appendix II and in the published
and unpublished literature.\cite{zachar-KLL,affleck,zachar}

\subsubsection{The decoupled Luttinger liquid: $J_{H}\gg -J_{K}\ge 0$}

As discussed above, spin-rotation invariance implies that, to lowest order,
the spin coupling between two gapless systems is marginal, while to second
order, as pointed out by Sikkema {\it et al},\cite{Affleck-zigzag} the
interactions are perturbatively irrelevant for $J_{K}<0$ (ferromagnetic
interactions) and relevant for $J_{K}>0$. Thus, for ferromagnetic Kondo
coupling, the decoupled fixed point is perturbatively stable. This phase
has, trivially, one gapless spin excitation at momentum $2k_{F}^{Heis}$, and
gapless spin and charge excitations at momentum $2k_{F}$. Since
$2k_{F}>2k_{F}^{Heis}$, this phase is labelled 
$[c,s;2k_{F}][s;2k_{F}^{Heis}]$,
and has no discrete broken symmetries. In short-hand notation, the
decoupled fixed point is classified as $[c,s][s]$.  
However, for future reference, it is important to note
that there is also a gapless, odd parity charge 2e
composite pairing excitation at $2k_F^{Heis}$, as discussed
in Appendix B.  Because this is a composite excitation, its existence 
is already implied by the existence of the other gapless 
modes.  However, as we shall now show,
when the Kondo coupling produces a spin gap, 
this composite mode can still remain gapless. 

\subsubsection{Antiferromagnetic Kondo coupling: $J_{K}>0$}

\paragraph{Odd parity singlet pairing: $J_{H}\gg J_{K}>0$.}

If $J_{K}>0$, the decoupled fixed point is unstable, and the low-energy
physics is governed by a strong-coupling fixed point with a spin gap.
\cite{Affleck-zigzag} This phase has several unexpected features. Of course, 
as required by the generalized Luttinger's theorem, there is a neutral,
spinless gapless excitation of this system with a minimal momentum $%
2k_{F}^{*}$, as pointed out by Yamanaka {\it et al}.\cite{theorem}
Remarkably the effective Fermi sea knows about the
localized electrons as well as the itinerant ones. In addition, this system
clearly has a gapless spinless charge 2e excitation created by the $\eta $%
-pairing operator of the 1DEG; in this case, since the localized spins are
unaffected by a global gauge transformation, our theorem implies that this
mode carries momentum $2k_{F}$. Of course these two statements, taken
together imply that there exists a gapless, spinless charge 2e excitation
with momentum $2k_{F}^{Heis}$. Because $4k_{F}^{Heis}=2\pi /b$ is a
reciprocal lattice vector, $2k_{F}^{Heis}$ and $-2k_{F}^{Heis}$ are
equivalent; as a consequence, excitations with momentum $2k_{F}^{Heis}$ can
simultaneously be characterized by their parity.  As discussed in Appendix B,
it may be shown that, in the present case, the only gapless charge 2e
excitation with momentum $2k_{F}^{Heis}$ has odd parity.\cite{Kondo-phases}
This phase is labelled $[c][c]$, or more fully as $%
[c;2k_{F}][(c,odd);2k_{F}^{Heis}]$ (where we put the modes in parentheses
when an additional descriptive element, such as even/odd parity, must be
noted).

There are some remarkable features of the charge fluctuations in this state. 
\cite{Kondo-phases} Whereas the decoupled system had gapless CDW modes at $2k_F$
and $2k_F^{\ast} = 2k_F + 2k_F^{Heis}$, the spin gap phase retains only the 
composite CDW mode at $2k_F^{\ast}$. (See Appendix B.) As a result, 
because of the mismatch between $2k_{F}^{Heis}$ and 
$2k_{F}$, there are no spinless charge 2e gapless excitations at momentum zero,
and the system cannot be a conventional superconductor with pairing at total
momentum zero induced by spin fluctuation exchange. An intuitive feeling for
the origin of the spin gap can be obtained by considering 
the strong-coupling limit of the model, although
care is always needed in identifying the specific strong-coupling fixed point to 
which a given weak coupling Hamiltonian flows. In the present case,
the natural candidate is a model
in which the 
conduction electrons form singlets with the localized spins, and any remaining 
localized spins form singlets with each other.\cite{Affleck-zigzag} 

\paragraph{Staggered pairing: $J_{K}\gg J_{H}>0$.}

An exact solution of the model\cite{zachar-KLL}, also with a spin gap, may
be obtained from a field theory with explicitly broken spin-rotational
symmetry and in a special ``Toulouse limit,'' in which one component of the
Kondo coupling takes a specific value $\sim E_{F}$. The renormalization
group strategy behind this solution was described in the introduction. The
long-distance behavior of the system in the Toulouse limit is
spin-rotationally invariant,\cite{alten,Col-Tsv-Geo} which implies that
spin-rotational symmetry breaking terms are irrelevant at the fixed point,
so it is unlikely that the behavior we found is an artifact of the model.
This phase may be distinguished\cite{Kondo-phases} from the weak-coupling ($%
J_{H}\gg J_{K}$) spin-gap state by classifying its gapless excitations.

Clearly, as before, there must exist a gapless neutral excitation with a
minimal momentum $2k_{F}^{*}$, a gapless charge 2e excitation, produced by
the $\eta $-pairing operator, with momentum $2k_{F}$, and as a consequence
of these two general statements, a gapless charge 2e excitation at momentum $%
2k_{F}^{Heis}$. However, we find\cite{zachar-KLL} that there exist both even
and odd parity gapless charge 2e excitations at momentum $2k_{F}^{Heis}$, so 
that there is, in fact, one more finite momentum gapless charge mode 
in this state than in the weak
coupling spin-gap state. This phase is labelled $[c][cc]$, or more fully as $%
[c;2k_{F}][(c,odd)(c,even);2k_{F}^{Heis}]$.
This solution also provides
an example of the fact that an analysis of the relevant operators at weak
coupling does not necessarily tell us the character of the stable fixed
point. It is long-distance physics (forward scattering) that destabilizes
the weak-coupling fixed point, but the character of the strong-coupling
fixed point is determined by short-distance physics.

An elaborate comparison\cite{Kondo-phases} of the two spin-gap phases
reveals that the additional gapless mode in the Toulouse limit phase may be
associated with an additional hidden broken translation symmetry. A further
distinction between the states may be made by considering the origin of the
spin gap and its consequences for enhanced pairing correlations;  an 
intuitive, strong-coupling picture of the origin of the spin gap 
in the Toulouse limit phase
involves pairing of the spins in each subsystem, separately.
The existence of the two spin-gap fixed points of the one dimensional
Kondo-Heisenberg model underscores the need to consider the explicit
solutions of the strong coupling fixed points, which do not follow from
simply establishing the existence of a spin gap based on the weak coupling
perturbative renormalization group analysis.

\paragraph{Strong coupling: $J_{K}\gg E_F \gg J_H$.}

Finally, for completeness, it is important to remark that 
there exists a direct ``strong-coupling'' limit of the model
$J_{K}\gg E_{F} \gg J_{H}$, which has distinct physics from either of the 
spin-gap phases discussed above.  In particular, via a mapping to 
the $t-J$ model, it has been shown that in this limit the system is 
governed by a Luttinger liquid fixed point with no 
spin gap.\cite{Z-staggered-phases,Sigrist(Kondo-strong)}
(Note, if $J_{H}\sim E_{F}$ there is also the possibility of a strong
coupling spin gap phase.\cite{Affleck-zigzag})

{\bf Aknowledgements:} We are grateful to I. Affleck and S. Sondhi for
discussions and comments. This work was supported, in part, by the NSF under
Grant No. DMR93-12606 at UCLA (S.A.K and O.Z.), by the Department of Energy
under Contract No. DE-AC02-98CH10886 (V.J.E.), and by the Chateaubriand
Fellowship and  TMR\#ERB4001GT97294 Fellowship (O.Z.).

\appendix

\section{The multicomponent free boson fixed point}

This appendix gives a derivation of Eqs. (\ref{eq:w}), (\ref{eq:v}), (\ref
{eq:delta}), (\ref{eq:m}), and (\ref{eq:n}). First of all, Eq. (\ref{eq:wv})
may be rewritten as the eigenvalue equation of a real symmetric matrix ${\bf %
W}_c^{{\frac{1 }{2}}}{\bf V}_c {\bf W}_c^{{\frac{1 }{2}}}$, with eigenvalues 
$u_{ci}^2$ and eigenvectors ${\bf W}_c^{{\frac{1 }{2}}} \hskip 0.1 cm {%
\underline \eta}_i$. It follows that 
\begin{equation}
{\bf W}_c^{{\frac{1 }{2}}} {\bf V}_c {\bf W}_c^{{\frac{1 }{2}}} = {\bf W}%
_c^{-{\frac{1 }{2}}} \sum_i u_{ci}^2 \hskip 0.1 cm {\underline \eta}_i %
\hskip 0.1 cm {\underline \eta}_i^T \hskip 0.1 cm {\bf W}_c^{-{\frac{1 }{2}}%
}.  \label{eq:symm}
\end{equation}
Equation (\ref{eq:w}) may now be obtained by premultiplying this equation by 
${\bf W}_c^{{\frac{1 }{2}}}$ and postmultiplying by ${\bf W}_c^{-{\frac{1 }{2%
}}}{\bf V}_c^{-1}$, and using Eq. (\ref{eq:wv}). Similarly Eq. (\ref{eq:v})
may be obtained by premultiplying and postmultiplying Eq. (\ref{eq:symm}) by 
${\bf W}_c^{-{\frac{1 }{2}}}$.

To derive Eq. (\ref{eq:delta}), first use Eq. (\ref{eq:w}) to rewrite Eqs. (%
\ref{eq:bogo}) as: 
\begin{equation}
{\underline a}^T {\underline {\tilde \phi}} = \sum_i u_i^{-{\frac{1 }{2}}}
\phi_{ci}^{\prime} \hskip 0.1 cm {\underline a}^T {\underline \eta}_i
\end{equation}
and 
\begin{equation}
{\underline b}^T {\underline {\tilde \theta}} = u_i^{{\frac{1 }{2}}}
\theta_{ci}^{\prime} \hskip 0.1 cm {\underline b}^T {\bf W}_c^{-1} {%
\underline \eta}_i
\end{equation}
Then, if Eq. (\ref{eq:operator}) is written as 
\begin{equation}
\hat O(x) = \sum_i (f_i \phi_i^{\prime} + g_i \theta_i^{\prime})
\end{equation}
where 
\begin{equation}
f_i = u_i^{-{\frac{1 }{2}}} {\underline a}^T {\underline \eta}_i
\end{equation}
and 
\begin{equation}
g_i = u_i^{{\frac{1 }{2}}} {\underline b}^T {\bf W}_c^{-1} {\underline \eta}%
_i
\end{equation}
the critical exponent $\delta$ is given by 
\begin{equation}
\delta = {\frac{1 }{4 \pi}} \sum_i (f_i^2 + g_i^2)
\end{equation}
from which Eq. (\ref{eq:delta}) follows with the aid of Eqs. (\ref{eq:vec})
and (\ref{eq:w}), and 
\begin{equation}
M = {\bf W}_c^{-1} \sum_i u_i {\underline \eta}_i {\underline \eta}_i^T %
\hskip 0.1 cm {\bf W}_c^{-1}.
\end{equation}
This equation may be rearranged to give Eqs. (\ref{eq:m}) and (\ref{eq:n})
by using Eq. (\ref{eq:symm}).

\section{Gapless modes of the Decoupled Kondo-Heisenberg array}

The study of the different stable phases of the 
Kondo-Heisenberg array begins with an analysis of the gapless 
excitations of 
the decoupled fixed point.  From there, as usual, we sort the phases 
by determining which of these excitations become gapped, and which 
remain gapless in the presence of the (Kondo) couplings between
the 1DEG and the Heisenberg chain.  
In this appendix, then, we analyze the system in
the absense of any Kondo interactions.  However, since our ultimate 
goal is to study the coupled system, we will consider the character 
of gapless excitations constructed of composites of the gapless modes 
of the two subsystems, as well as the excitations of 
each, separate subsystem.

The  low energy spin currents of the 1DEG, $\vec{s}\left( x\right) $, can be
decomposed into two parts; 
\begin{equation}
\vec{s}\left( x\right) =\vec{J}_{1}\left( x\right) +\left [ \vec{n}_{1}\left(
x\right) e^{i2k_{F}x} + {\rm H.c.}\right] \nonumber
\end{equation}
where $\vec{J}_{1}=\frac{1}{2}\sum_{\lambda,\sigma ,\sigma ^{\prime
}}\psi _{\lambda,\sigma }^{\dagger }\vec{\sigma}_{\sigma ,\sigma ^{{\prime }%
}}\psi _{\lambda,\sigma ^{\prime }}$ and $\vec{n}_{1}=\frac{1}{2}%
\sum_{,\sigma ,\sigma ^{\prime }}\psi _{1 ,\sigma }^{\dagger }%
\vec{\sigma}_{\sigma ,\sigma ^{{\prime }}}\psi _{-1,\sigma ^{\prime
}} $ are respectively the $k=0$ and the $k=2k_{F}$ components of the SDW
(charge-$0$, spin-$1$) of the 1DEG.

The Heisenberg chain spin current, $\vec{\tau}_{j}$, may be similarly decomposed
into a $k=0$ part, $\vec{J}_{\tau }$, and a finite momentum $k=\frac{\pi }{b}
$ part, $\left( -1\right) ^{j}\vec{n}_{\tau }$ (where $\frac{2\pi }{b}$ is
the reciprocal lattice vector of the Heisenberg chain); 
\begin{equation}
\vec{\tau}_{j}=\vec{J}_{\tau }\left( x_{j}\right) +\left( -1\right) ^{j}\vec{%
n}_{\tau }\left( x_{j}\right)
\end{equation}

As explained in the introduction, we count only the number of finite
momentum excitations. It follows from time reversal 
symmetry that, for finite momentum, if there
is a gapless mode at momentum $q$ than there is also a gapless mode with
momentum $-q$. We count them as one mode. To summarize, the 
gapless spin-1 excitations of
the 1DEG and the Heisenberg spin chain, and the operator whose 
correlation function is most directly sensitive to it are listed in table-B-I. 
\begin{equation}
\stackrel{Table-B-I:\text{ Gapless SDW excitations}}{
\begin{tabular}{||c|c||}
\hline\hline
$
\begin{array}{c}
operator
\end{array}
$ & $
\begin{array}{c}
wave \\ 
number
\end{array}
$ \\ \hline
$\vec{n}_{1}$ & $2k_{F}$ \\ \hline
$\vec{n}_{\tau }$ & $\frac{\pi }{b}$ \\ \hline\hline
\end{tabular}
}
\end{equation}

The incommensurate 1DEG has one charge-$0$ spin-$0$ CDW excitation 
with momentum, $2k_{F}$, created by the operator 
\begin{equation}
\hat{O}_{CDW}=\frac{1}{2}\sum_{\lambda ,\sigma }\psi _{\lambda ,\sigma
}^{\dagger }\psi _{-\lambda ,\sigma }. 
\end{equation}
The reader may notice an 
apparent
conflict with the ``prediction'' of the generalized Luttinger theorem that
there must be a gapless CDW mode at $2k_{F}^{*}=2k_{F}+\frac{\pi }{b}$. This
conflict is resolved by realizing the existence of composite-CDW\cite
{Kondo-phases,Col-Tsv-Geo} order parameters which are formed by combining a
spin-$1$ SDW of the 1DEG with a spin-$1$ SDW of the Heisenberg chain into a
composite singlet $\hat{O}_{c-CDW}$, 
\begin{eqnarray}
\hat{O}_{c-CDW}&=& \vec{s}\cdot \vec{\tau} \nonumber \\
&=&\vec{J}_{1}\cdot \vec{J}_{\tau }+\vec{J}_{1}\cdot \vec{n}_{\tau}
\left( -1\right) ^{j}  \nonumber \\
&+&\left[\vec{n}_{1}\cdot \vec{J}_{\tau }e^{i2k_{F}x} 
+ H.c.\right] \nonumber \\
&+&\left[\vec{n}_{1}\cdot \vec{n}_{\tau }e^{i2k_{F}x} + H.c. \right]
\left( -1\right) ^{j}. 
\end{eqnarray}
The staggered component, $\vec{n}\cdot \vec{n}_{\tau }$,
has momentum $2k_{F}^{*}=2k_{F}+\frac{\pi }{b}$, and is
thus the CDW excitation required by the generalized Luttinger theorem. There is
also a new composite-CDW excitation $\vec{J}\cdot 
\vec{n}_{\tau }$
at $k=\frac{\pi }{b}$. To summarize, the non-interacting
two-chain system of a Luttinger liquid and a Heisenberg spin chain has {\em %
gapless CDW modes at three wave vectors} (table-B-II). 
\begin{equation}
\stackrel{Table-B-II:\text{ Gapless CDW excitations}}{
\begin{tabular}{||c|c||}
\hline\hline
$
\begin{array}{c}
operator
\end{array}
$ & $
\begin{array}{c}
wave \\ 
number
\end{array}
$ \\ \hline
$\vec{n}_{1}\cdot \vec{n}_{\tau }$ & $2k_{F}+\frac{%
\pi }{b}$ \\ \hline
$\hat{O}_{CDW}$ & $2k_{F}$ \\ \hline
$\vec{J}_{1}\cdot \vec{n}_{\tau }$ & $\frac{\pi }{b}$
\\ \hline\hline
\end{tabular}
}
\end{equation}
Note that the composite-CDW excitations at wave vectors $\frac{\pi }{b}$ and $%
2k_{F}+\frac{\pi }{b}$ are not independent, since they can be related
through a multiplication by the 1DEG $\hat{O}_{CDW}$ (which has wave vector $%
2k_{F}$).

The charge-$2e$ singlet pairing modes also require careful consideration. In
addition to the usual $k=0$ BCS even parity singlet pairing, 
\[
\Delta =\frac{1}{\sqrt{2}}\sum_{\lambda }\psi _{\lambda ,\uparrow }^{\dagger
}\psi _{-\lambda ,\downarrow }^{\dagger }, 
\]
we note also the existence of an $\eta $-pairing mode, $\psi _{\lambda
,\uparrow }^{\dagger }\psi _{\lambda ,\downarrow }^{\dagger }$, at momentum $%
\pm 2k_{F}$ , corresponding to right and left going singlet pairs.

As with the CDW modes, in addition to the singlet pairing modes of the 1DEG,
it is necessary to consider the {\em composite} singlet pairing, $O_{c-SP}$,
(a product of a triplet pairing in the 1DEG with a spin-$1$ mode of the
Heisenberg chain) which turns out to be odd parity\cite
{odd-w,Coleman-Miranda}, 
\begin{equation}
O_{c-SP}=-i\frac{1}{2}\left( \psi _{1}^{\dagger }{\bf \vec{\sigma}}\sigma
_{2}\psi _{2}^{\dagger }\right) \cdot {\bf \vec{\tau}}
\end{equation}
where the sum over the spin-indices of the spinors is implicit.
It is decomposed into two momentum components:  a uniform $k=0$ composite odd
parity singlet 
\begin{equation}
\hat{O}_{c-SP}^{k=0}\left( x\right) =-i\frac{1}{2}\left( \psi _{1}^{\dagger }%
{\bf \vec{\sigma}}\sigma _{2}\psi _{2}^{\dagger }\right) \cdot \vec{J}_{\tau
}  \label{eq:k=0 c-sp}
\end{equation}
and a $k=\frac{\pi }{b}$ , {\it i.e.} a {\sl staggered} composite odd parity singlet 
\begin{equation}
\hat{O}_{c-SP}^{stagger}\left( x\right) =-i\frac{1}{2}\left( \psi
_{1}^{\dagger }{\bf \vec{\sigma}}\sigma _{2}\psi _{2}^{\dagger }\right)
\cdot \vec{n}_{\tau }\left( -1\right) ^{j}  \label{eq:staggered- c-sp}
\end{equation}

The pairing and CDW modes can be related through the $\eta $-pairing modes
by using the identities\cite{Kondo-phases}: 
\begin{eqnarray}
\left[ \hat{O}_{CDW},\eta ^{even}\right] &=&\Delta
\label{eq:pairing-commute1} \\
\left[ \hat{O}_{c-CDW},\eta ^{odd}\right] &=&\hat{O}_{c-SP}.
\end{eqnarray}
where, 
\begin{eqnarray}
\eta ^{even} &\equiv &\frac{1}{\sqrt{2}}\sum_{\lambda =\pm }\psi _{\lambda
,\uparrow }^{\dagger }\psi _{\lambda ,\downarrow }^{\dagger } \\
\eta ^{odd} &\equiv &\frac{1}{\sqrt{2}}\sum_{\lambda =\pm }\lambda \psi
_{\lambda ,\uparrow }^{\dagger }\psi _{\lambda ,\downarrow }^{\dagger }.
\end{eqnarray}
The above relations are a manifestation of the fact that for each gapless
pairing mode there is a corresponding gapless CDW mode. Therefore, we adopt
the custom of dropping the CDW modes from the explicit notation, (but their
``trivial'' existence should be implicitly understood in any fixed point
which has the corresponding gapless charge-2e gapless mode).

In summary, there are two independent gapless, finite momentum, charge-$2e$
pairing modes.
\begin{equation}
\stackrel{Table-B-III:\text{ Gapless charge-2e pairing excitations}}{
\begin{tabular}{||c|c||}
\hline\hline
$
\begin{array}{c}
operator
\end{array}
$ & $
\begin{array}{c}
wave \\ 
number
\end{array}
$ \\ \hline
$\eta $ & $2k_{F}$ \\ \hline
$\hat{O}_{c-SP}$ & $\frac{\pi }{b}$ \\ \hline\hline
\end{tabular}
}
\end{equation}

The gapless modes in tables I-III characterize the non-interacting fixed
point of a two-chain system consisting of a 1DEG Luttinger liquid with Fermi
wave number $2k_{F}$ and a Heisenberg spin chain with a reciprocal lattice
vector $\frac{\pi }{b}$. In our compact notation, which counts only the
singlet charge-2e and spin-1 charge-0 modes at finite momentum, 
the decoupled fixed point is denoted by 
\begin{equation}
\lbrack c,s;2k_{F}][s;\frac{\pi }{b}]
\end{equation}
or $[c,s][s]$ for short.  However, as emphasized in the above, this 
description leaves implicit the gapless, odd parity charged 
excitation at momentum $\pi/b$.  Since this excitation is a composite 
of the excitations already listed, it can be omitted
in a minimal labelling 
scheme.  But because of this, it looks a bit mysterious that there
appears a a gapless composite pairing mode with momentum $\pi/b$ 
in the spin-gap states which appear under the influence of a relevant 
perturbation - it looks (incorrectly) as if the relevant interaction 
is generating new gapless excitations.  From 
this viewpoint, one might be tempted to label the decoupled fixed 
point 
$[c][c,s]$, and to view the spin one excitation of the 1DEG as a 
composite of the three other modes, however far that is from the 
actual physics of two decoupled systems.  The non-uniqueness of 
the label associated with each state is an intrinsic feature of the 
approach taken in the present paper.


\begin{references}
\bibitem{spinmodes}  Note that if there exist gapless spin-n modes with
wave-vectors $\pm 2k_{F}$ and any integer $n>0$, it follows that one can
construct gapless states out of multiple excitations of these gapless modes
with any integer spin. Thus, the case we have considered here, in which the
spin mode has spin-1, is in fact completely general. There are, however,
some exotic possibilities that we have not considered here explicitly, as
they do not seem to occur in any model problems of which we are aware. 
For example, it is conceivable that the fundamental gapless mode is spin 1, 
but charge 2e.
\bibitem{pair}  If there exists a gapless, spinless charge $2e$ mode, one
can clearly construct gapless, spinless excitation with any charge $2ne$.
One can imagine cases in which there are no gapless charge $2e$ excitations,
but there are gapless spinless excitations with charge $4e$ and multiples
thereof. We will adopt the convention that the fundamental charged, spinless
excitation is assumed to have charge $2e$, unless otherwise specified.
\bibitem{stuff}  Where there is a gapless spin mode, a CDW fluctuation can
be viewed as a total spin-zero composite of three spin modes. Where there is a
spin gap but no charge gap, the generalized Luttinger's theorem implies that
there must be a gapless, zero momentum charge $2e$ ``BCS mode'', which can
be thought of as a composite of an $\eta $-pairing mode and a CDW mode;
conversely, the CDW mode can be thought of as a composite of oppositely
charged $\eta $-pairing and BCS modes. This ambiguity is a reflection of the
number-phase duality of the charge modes.
\bibitem{gap}  Here, we have considered only the case in which a gap 
results from a relevant interaction that pins the
value of the density-wave phase, $\phi$.  While it is typically 
forbidden by symmetry in most cases, it is certainly possible to 
imagine cases in which there is a relevant interaction of the form 
$\cos(\beta^{\prime}\theta_{\alpha})$ which would pin the dual phase;  
we do not explore this possibility in the present paper.
\end{references}
\end{document}